\documentclass[10pt,a4paper]{article}
\usepackage{graphics}
\usepackage{amssymb}
\begin{document}

\title{Calculating asymptotic quantities near space-like and null infinity
from  Cauchy data} 

\author{Helmut Friedrich\\
Albert-Einstein-Institut\\
Max-Planck-Institut f\"ur Gravitationsphysik\\
Am M\"uhlenberg 5\\
14476 Golm, Germany\\ \\ 
J\'anos K\'ann\'ar\\
MTA KFKI \\
R\'eszecske \'es Magfizikai Kutat\'oint\'ezet \\
Budapest, Pf. 49, \\
1525, Hungary}

\date{}

\maketitle

\begin{abstract}
Following \cite{FrKan}, we discuss how asymptotic quantities,
originally introduced on null infinity in terms of Bondi-type gauge
conditions, can be calculated near space-like infinity to any desired
precision.
\end{abstract}

\section{Introduction}

Null infinity is the most interesting asymptotic domain of asymptotically flat
space-times, since it is here where information on dynamical processes which
take place in the interior of space-times is registered by the ideal observer
at infinity. The early investigations of asymptotical flatness in null
directions (\cite{BBM}, \cite{NP:spin-coefficients}, \cite{S}) are based on
the notion of ``Bondi-type'' coordinates and associated gauge conditions,
which involve as a basic ingredient families of outgoing null hypersurfaces
of the form $\{u = const.\}$ with a smooth function $u$. Such coordinates play
a distinguished role in these studies, because the fall-off behavior of the
fields at null infinity, put in after a judicious choice ``by hand'', is
specified in terms of suitable parameters along the generators of the null
hypersurfaces. 

In the following we shall mainly be concerned with the idea of asymptotic
simplicity (\cite{P:asymptotic-properties-of-fields},
\cite{P:zero-rest-mass-fields}). This proposal to characterize the fall-off
behavior of gravitational fields in terms of the light cone structure does
not require the use of preferred classes of coordinates. Nevertheless,
because null hypersurfaces are  conformally invariant structures and are
closely related to certain features of gravitational field propagation,
Bondi-type gauge conditions have been used so extensively also in this
context that they may appear to be naturally associated with the analysis of
asymptotic properties of gravitational fields. We shall see in the following
that there are good reasons to consider also other coordinates and gauge
conditions in the study of asymptotics. 

A detailed understanding of the behavior of gravitational fields near
space-like infinity plays a key role in the analysis of asymptotically flat
space-times. As pointed out already in \cite{P:zero-rest-mass-fields}, such an
understanding is required to relate information given on past null infinity
to information gathered on future null infinity. Moreover, it is obviously
desirable to complete the interpretational frame-work developed for the
gravitational fields of isolated systems by relating concepts like
``radiation field'', ``Bondi-energy-momentum'', etc. which have been
introduced on null infinity, to concepts defined at space-like infinity.
Other reasons for studying space-like infinity will be seen below. Several
attempts have been made in this direction. In \cite{ashtekar II} 
(cf. also the references given in this article), \cite{geroch} this
was done on the level of conformal geometry without touching the field
equations in an essential way. Thus there arose the need to introduce further
assumptions on top of to those made already in the study of null infinity.
Later approaches (\cite{beig}, \cite{beig:schmidt}), which have
subsequently been transposed into the conformal picture
\cite{ashtekar:romano}, took the field equations to some extent into account
but they offered no obvious way to relate information near space-like
infinity to structures on null infinity. Moreover, it appears difficult to
bring these studies in relation to those of the Cauchy problem, something
which one would like to do to replace implicit assumptions on the evolution
in time by definite statements about sufficiently general classes of
solutions with required properties. 

The work \cite{christ:klai} contains a discussion of space-like infinity
which does not involve any assumptions besides those on initial data and
which, in a sense, is rather complete. However, for reasons indicated in the
introduction of \cite{christ:klai}, no attempt has been made to clarify the
notion of asymptotic simplicity. Earlier studies of asymptotic simplicity in
terms of the conformal field equations (cf. the references given in
\cite{Fr:conformal-structure-survay}, \cite{Fr:GR15}) have shown that, at
least for solutions to the vacuum field equations arising from small initial
data, the smooth extensibility of the light cone structure through null
infinity postulated by asymptotic simplicity is preserved along null
infinity if it holds close to space-like infinity. In this sense the
decision whether there exists a smooth structure at null infinity is made at
space-like infinity.  

Whatever techniques are used to analyze the field near space-like infinity,
the slow fall-off behavior of the metric field near space-like infinity
will give rise to difficulties. In terms of the conformal picture the latter
take the following form. In the standard conformal completion of Minkowski
space space-like infinity is represented by a point $i^0$. Any smooth
space-like Cauchy hypersurface $\tilde{S}$ of Minkowski space approaches this
point. To $\tilde{S}$ can be adjoined a point $i$ to obtain a $3$-sphere 
$S = \tilde{S} \cup \{i\}$ such that $i$ will coincide with $i^0$ if the
immersion of $\tilde{S}$ into compactified Minkowski space is extended to
cover $S$. We shall consider the case where $\tilde{S}$ is chosen such that
$S$ will be smoothly embedded. Then the conformal initial data induced on
$S$ will be smooth and the conformal Weyl tensor will vanish. If these data
are now deformed while keeping the conformal vacuum constraints being
satisfied, they will loose their smoothness at $i$, in fact, the conformal
Weyl tensor will even diverge there, as soon as the ADM mass of the data set
does not vanish any longer. This fact, a reflection of the slow fall-off of
the metric, represents the technical hitch which obstructs the immediate
application of the conformal field equations near space-like infinity. 

In \cite{Fr:space-like-infinity} has been shown that under suitable
assumptions on the data some of these difficulties can be overcome. It
turns out that it is possible to introduce a setting, defined in terms
of conformally invariant structures, in which the asymptotic regions
are represented in a finite picture, the data extend smoothly to a
suitable defined ``space-like infinity'', and in which the equations
are rendered in a manageable form. Moreover, the setting allows us to
relate near space-like infinity structures at null infinity to the
data on the initial hypersurfaces. Some of the consequences of this
setting will be recalled in chapter 2. The gauge conditions employed
in this study are completely different from Bondi-type gauge
conditions. Therefore we have analysed in \cite{FrKan} to what extent
the discussion of null infinity in terms of Bondi-type systems can be
redone near space-like infinity in the setting of
\cite{Fr:space-like-infinity}. As discussed in chapter 3, the
transformation which relates the different gauge conditions becomes
singular at the set where ``null infinity touches space-like
infinity'' (cf. the sets $I^{\pm}$ below). However, it does so in a very
controlled way and it actually allows us to calculate near space-like
infinity expansions of the quantities of physical interest, such as the
radiation field on null infinity, in terms of the initial data. As a
particular application we have determined a formula for the NP-constants in
terms of the initial data (chapter 4). In the course of these calculations
we have also determined a certain expansion near space-like infinity 
introduced in \cite{Fr:space-like-infinity} to third order.
The regularity of the third order expansion coefficient sheds some light on
the possible smoothness at null infinity under our assumptions on the
fall-off behavior of the data (chapter 4). In conclusion it can be said
that, apart from the length of the calculations involved, Bondi-type systems
can conveniently be related to the setting of \cite{Fr:space-like-infinity}
and that our investigation sheds light on the analysis of 
\cite{Fr:space-like-infinity} as well as on the investigations in terms of
the Bondi-type gauge conditions.

\section{The regular finite initial value problem near space-like infinity}

It turns out that some of the difficulties near space-like infinity
indicated above can be circumvented by using a new representation of
the conformal field equations \cite{Fr:AdS} and a suitable choice of gauge
conditions \cite{Fr:space-like-infinity}. In the resulting finite picture
space-like infinity is represented by a cylinder $I = I^0 \times ]-1,
1[$ instead of by a point $i^0$. The initial data for the conformal
field equations are given on a manifold with boundary, where the latter is
obtained by adjoining to $\tilde{S}$ the spherical set $I^0 \times \{0\}
\simeq I^0$ instead of the point $i$. The initial data for the conformal
field equations (derived from asymptotically Euclidean vacuum data)
extend smoothly to the boundary of this manifold, and the conformal
field equations imply in this setting symmetric hyperbolic evolution
equations for the extended data. The gauge conditions underlying this
picture are again defined in terms of conformally invariant structures
but they are completely different from the Bondi-type gauge conditions
referred to above.  Nevertheless, if the solution of the initial value
problem so obtained extends near space-like infinity with sufficient
smoothness sufficiently far into the future, we will be able to
identify near $I$ in our coordinates future and past null infinity as
null hypersurfaces ${\cal J}^{\pm}$ of a finite location which touch
the cylinder $I$ at its boundary components $I^{\pm} = I^0 \times
\{\pm 1\}$. In a suitable gauge (different from the one used in the
calculations referred to below) a neighborhood $M$ of space-like infinity,
which includes the asymptotic structures, can be described in the form $M =
\{(\tau, \rho, t) \in \mathbb{R} \times [0, a[ \times I^0\,| \,\,|\tau| \le 1
+ \rho\}$, with some number $a > 0$, such that the subsets of $M$
referred to above are then given by $\tilde{S} = \{\tau = 0, \rho > 0\}$, $I
= \{|\tau| < 1, \rho = 0\}$, $I^{\pm} = \{\tau = \pm 1, \rho = 0\}$, ${\cal
J}^{\pm} = \{\tau = \pm (1 + \rho), \rho > 0\}$, the ``physical part'' of
$M$ is given by $\{|\tau| < 1 + \rho, \rho > 0\}$.  We consider $\tau$ and
$\rho$ as functions on $M$. Near $\tilde{S} \cup I$ will then be defined on
$\tilde{M}$ a time function by $\tau$ and $\rho$ can be understood as a
radial coordinate which vanishes at space-like infinity.

In \cite{Fr:space-like-infinity} the picture outlined above has been
analyzed under the simplifying assumptions that the data are time-symmetric
and that the conformal structure induced on the initial hypersurface
$\tilde{S}$ extends smoothly (in fact analytically) to $i$, conditions which
have also been assumed in the article \cite{FrKan} on which we shall report
below. Since these conditions appear rather restrictive, we remark that
recent studies demonstrated possibilities for generalizing this
analysis to more general data \cite{dain}. The smooth conformal initial data
on $\tilde{S} \cup I^0$ define together with the conformal evolution
equations the {\it regular finite initial value problem near
space-like infinity}. This setting has been developed with the aim to derive
statements about the smoothness of the solutions near the sets
${\cal J}^{\pm} \cup  I^{\pm}$. To achieve this goal, we need to understand 
the time evolution near the set $I$ in sufficient detail. To some extent
this has been studied already in \cite{Fr:space-like-infinity}. We recall
some of the results below without going into details. The article
\cite{FrKan} is a further step into this direction.

Since the conformal data are given on $\tilde{S} \cup I^0$ and the
conformal field equations are to be solved on $M$, one may be led to think
that one has to solve a boundary value problem with time-like boundary $I$.
This is not correct. The unknown of the conformal field equations
can be written schematically $u = (frame,\;connection,\;curvature)$ and the
evolution equations take the form
\begin{equation}
\label{evol}
\{A^{\tau}\partial_{\tau} + A^{\rho}\,\partial_{\rho} + A\}\,u = B\,u,
\end{equation}
with a differential operator $A$ on $I^0$ and matrix-valued functions
$A^{\tau}$, $A^{\rho}$, $B$ which depend on the coordinates and the unknown.
The equations are symmetric hyperbolic in some neighborhood $N$ of the
initial hypersurface and the data determine a unique smooth solution on $N$.
Though we can imagine $I$ as arising as a limit of hypersurfaces in $N\cap
\tilde{M}$  which are time-like for the metric defined by the solution, the
metric, which is smooth and Lorentzian on $N \cap \tilde{M}$, degenerates as 
$\rho \rightarrow 0$ and ``time-like'' looses its meaning on $I$. This fact
is reflected by the observation that $A^{\rho} = 0$ on $I \cap N$.
Consequently, the equations above, if restricted to $I$, imply an interior
system on $I$ which determines $u$ on $I$ uniquely from the data $u|_{I^0}$,
i.e. there is no freedom to prescribe boundary data on $I$. Moreover,
applying repeatedly the operator $\partial_{\rho}$ to the equations and
restricting to $I$, we get interior equations for the quantities 
$u^p = \partial^p_{\rho}\,u|_{I}$ on $I$, which allow us to determine a
formal expansion  $u = \sum_{p \ge 0}u^p\,\frac{1}{p!}\,\rho^p$ of the
solution at $I$. Under the assumptions of \cite{Fr:space-like-infinity} this
series converges near $I^0$ and provides the solution. 

The calculation of the function $u^0$, which allows us to determine 
$A^{\tau}$ on $I$, leads to the following important observation: the
matrix $A^{\tau}$, which is positive definite on $I$, extends smoothly to
$I^{\pm}$ but drops rank there by $1$. Following the procedure outlined
below, the quantities $u^p$, $p = 0, 1, 2$, have been calculated explicitly in 
\cite{Fr:space-like-infinity}. Using them to determine
$\partial_{\rho}\,A^{\tau}$ on $I$ we get an expression which suggests that
the degeneracy of $A^{\tau}$ is removed as soon as $\rho > 0$. 
{\it Thus all the remaining problems to analyze space-like infinity can be
traced back to the degeneracy of $A^{\tau}$ at $I^{\pm}$}.

Expanding $u^p$ in terms of a suitable system of functions
$T_m\,^j\,_k$ on the sphere $I^0$ (related to spin spherical
harmonics), the interior equations are reduced to systems of ODE's for
the $\tau$-dependent expansion coefficients. The task of solving these
equations can be reduced to solving a hierarchy of ODE's of the form
$\frac{d}{d\tau}y = C\,y + b$, with $2 \times 2$-matrix-valued
functions $C$ and $2$-component vector-valued function $b$ and $y$ of
$\tau$. The unknowns $y = y(\tau)$ as well as the coefficients of the systems
carry a ``multi-index'' $\alpha = (p, m, j, k)$ which has been suppressed
here. The components of $y$ are related to derivatives, whose order
depends on $p$, of certain components of the rescaled conformal Weyl
tensor. For given order $p$ the functions $b$ are calculated (essentially) from the
$u^q$, $q \le p - 1$. The matrices $C$ are universal in the sense that
they depend on the number $p$ but not on the quantities $u^q$, $q = 1,
2, \ldots$ The solutions to these equations can be written in the form
\begin{equation}
\label{gensol}
y(\tau) = X(\tau)\,X(0)^{-1}\,y_0
+  X(\tau)\,\int_0^{\tau} X(\tau')^{-1}\,b(\tau')\,d\tau',
\end{equation}
where $y_0 = y(0)$ and $X(\tau)$ (suppressing again the multi-index $\alpha$)
denotes a fundamental matrix of the given ODE. In
\cite{Fr:space-like-infinity} these fundamental systems have been given
explicitly in terms of Jacobi polynomials. Thus the solutions can be
calculated recursively. Since for increasing $p$ the explicit expressions
for $b$ become quite complicated the solutions $y(\tau)$ have not been
analyzed in all details yet. 

However, we can say the following. Due to the degeneracy of $A^{\tau}$ on
$I$ the ODE's for the $y$'s are singular at $\tau = \pm 1$. As a consequence,
we find for any $p \ge 2$ certain multi-indices $\alpha^*$ for which the
fundamental system $X$ develops logarithmic singularities at $\tau = \pm 1$.
Since the corresponding functions $b$ vanish, there cannot arise
cancellations from the integrals in (\ref{gensol}). {\it Thus, in general, the
functions $u^p$ develop such singularities for all $p \ge 2$ on $I^{\pm}$}. 

Provided the solution of our initial value problem extends to the sets
denoted above by ${\cal J}^{\pm}$ and the metric remains sufficiently
smooth there, these sets may still be null hypersurfaces for the solution,
whence characteristics for the evolution equations. The hyperbolicity of the
conformal evolution equations then suggests that the logarithmic
singularities which develop at $I^{\pm}$ will be transported along the
generators of ${\cal J}^{\pm}$. Several authors speculated about the
occurrence of logarithmic singularities at null infinity. The observation
above provides perhaps the clearest support for this view. This picture
would also be consistent with the results of \cite{christ:klai}, where
estimates for the conformal Weyl curvature have been obtained which suggest
a peeling which is weaker than that implied by asymptotic simplicity. It
should be noted that the logarithmic singularities observed here occur in
spite of the fact that we have chosen initial data which are particularly
``clean'' at space-like infinity.

Because of the degeneracy of $A^{\tau}$ at $I^{\pm}$ it is not clear that
the solution would extend smoothly to ${\cal J}^{\pm}$ near $I^{\pm}$ if all
the functions $u^p$ on $I$ would extend smoothly to $I^{\pm}$. Before
trying to solve this problem, we should consider another one: can we
provide situations, without being led back to Minkowski space, where the
$u^p$ extend smoothly to $I^{\pm}$ ? Other studies of asymptotic simplicity
(cf. the discussions in \cite{Fr:conformal-structure-survay}, \cite{Fr:GR15})
suggest that the logarithmic singularities might be avoided if the data are
subject to suitable extra conditions. Inspecting for the multi-indices
$\alpha^*$ the first (and in that case the only) term on the right hand
sides of (\ref{gensol}), we find after a somewhat lengthy analysis of the
initial data that the logarithmic singularities at $I^{\pm}$ observed above
do not occur if and only if the free data on $S$, represented under our
assumptions by the intrinsic conformal 3-metric $h$, satisfies the {\it
asymptotic regularity conditions}  
\begin{equation}
D_{(a_p b_p} \ldots D_{a_1 b_1}\,b_{abcd)}(i) = 0,
\label{asregcond}
\end{equation}
for $p = 0, 1, 2, \ldots$. Here we use space-spinor notation, the spinor
field $b_{abcd} = b_{(abcd)}$ represents the Cotton tensor of $h$, and $D$
denotes the covariant Levi-Civita derivative of $h$. If we want the $u^p$
to extend smoothly to $I^{\pm}$ only for $p \le p_*$, we need to require   
(\ref{asregcond}) only for $p \le p_* - 2$. We note that for given integer 
$p_* \ge 0$ the requirement that (\ref{asregcond}) hold for $p \le p_*$ is
conformally invariant and thus provides in fact a condition on the conformal
structure of $h$, i.e. on the free data. For the question of generality of
the class of metrics satisfying (\ref{asregcond}) we refer to 
\cite{Fr:space-like-infinity}.  

If (\ref{asregcond}) is not satisfied at some order, logarithmic terms would be
fed into some of the function $b$ on the right hand sides of (\ref{gensol}) and
result in logarithmic singularities at higher order. But even if
(\ref{asregcond}) holds at any order there may still arise difficulties. As a
further consequence of the degeneracy of $A^{\tau}$ on
$I^{\pm}$ we find that $\det(X) \rightarrow 0$ as $\tau \rightarrow \pm 1$.
Thus we might find singularities at $I^{\pm}$ in the solutions $y$ for
those multi-indices $\alpha$ (with $p \ge 3$) for which the function $b$ in
the integrals on the right hand sides of (\ref{gensol}) does not vanish.      
Since the occurrence of singularities depends on the form of $b$ it also
depends very much on the specific algebraic structure of the conformal
Einstein equations. We will say more about this below.

\section{Bondi-type gauge conditions near $I$}

In \cite{FrKan} we analyzed the relation of the early discussions of fields
near null infinity in terms of Bondi-type gauge conditions, in particular
the results of \cite{NP:NPQ-letter}, \cite{NP:NPQ}, to the setting provided in
\cite{Fr:space-like-infinity}. One of our main questions here was: 
do quantities and physical concepts introduced on ${\cal J}^+$ in the
NP-gauge extend in a meaningful way to $I^+$ in the F-gauge ?

The first problem is to relate the different gauge conditions involved. Given a
smooth structure on null infinity, Bondi-type gauge conditions can be fixed
conveniently near ${\cal J}^+$. One chooses a (spherical) space-like cut 
${\cal C}$ of ${\cal J}^+$, removes some of the gauge freedom on ${\cal C}$,
and removes further gauge freedom on ${\cal J}^+$ by solving certain ODE's
along the generators of ${\cal J}^+$. In the course of this is defined a
function $u$ on ${\cal J}^+$ which defines a space-like slicing of ${\cal
J}^+$ with $u = 0$ on ${\cal C}$. The function $u$ is extend into the physical
space-time such that the hypersurfaces $\{u = const.\}$ are null and some of
the remaining gauge conditions are fixed by some propagation law along the
generators of these hypersurfaces. There are variants of this. We consider in
\cite{FrKan} the gauge employed in \cite{NP:NPQ}, which involves a choice of
conformal factor, of coordinates, and of a frame field, and refer to it as to
the NP-gauge.   

The gauge conditions used in \cite{Fr:space-like-infinity}, which
involve similar fields, are determined by the initial data on $\tilde{S} \cup
I^0$ and certain propagation laws which are encoded in an implicit way into
the form of the conformal evolution equations. We refer to this gauge as to the
F-gauge. If we want to discuss the NP-gauge in this context we clearly have to
assume that the solution to our initial value problem extends smoothly to
${\cal J}^{\pm}$. In the detailed calculation we will in fact assume
that the solution extends smoothly to ${\cal J}^{\pm} \cup I^{\pm}$. We
impose, of course, the asymptotic regularity condition on the data.

The solution of the initial value problem is not given explicitly.
Consequently, the transformation from one gauge into the other, which involves
a conformal rescaling, a rotation of the frame, and coordinate transformation,
cannot be calculated explicitly. We therefore pursue the following strategy. We
describe in an abstract way the procedure to get from the F-gauge into the
NP-gauge. Here we use the a priori information given by the setting of the
regular finite initial value problem, i.e. the known location of ${\cal J}^+$
and the explicitly given conformal factor $\Theta$. We try to push the section
${\cal C}$ to $I^+$. Then we integrate the fields $u^p$ to the desired order
and determine an expansion of the transformation, respectively of certain
transformed fields, in terms the coordinates given by the F-gauge. The
feasibility of this procedure depends on the functions $u^p$ needed on $I^+$ 
and on the behavior of the ODE's on ${\cal J}^+$ near $I^+$ which need to
be solved to achieve the NP-gauge.

It turns out that after suitable redefinitions of the unknowns and the
equations we obtain equivalent ODE's on ${\cal J}^+$ which extend in a
regular way to $I^+$. Using $\rho$ as a parameter on the generators on 
${\cal J}^+$ we can describe the basic result on ${\cal J}^+$ near 
$I^+ = \{\rho = 0\}$ as follows.

(i) The positive function $\theta$ which we need to redefine the conformal
factor by $\Theta \rightarrow \theta\,\Theta$ is of the form
$\theta = 1 + O(\rho)$.

(ii) The transformation $\Lambda \in SL(2,C)$ relating the
spine frames is of the form 
\[
\Lambda =
\rho^{\frac{1}{2}}\,\left( \begin{array}{cc}
O(\rho) & 1 + O(\rho) \\
- \frac{1}{\rho} + O(\rho^0) & O(\rho^2)
\end{array} \right).
\]

(iii) The affine parameter on the generators of ${\cal J}^+$ in the NP-gauge
is of the form 
$u = \sqrt{2}\,\{- \frac{1}{\rho} + 4\,m\,\log \rho + u_* +  O(\rho)\}$. 

In these equations $u_*$ denotes a constant and an expression $O(\rho^k)$
stands for a smooth functions on ${\cal J}^+ \cup I^+$ whose Taylor expansion
at $\rho = 0$ starts with the power $k$. Obviously, $\theta$ extends
smoothly to $I^{\pm}$. The transformation $\Lambda$ does not extend smoothly
but has a well controlled behavior at $I^{\pm}$. If it is used e.g. to
transform the rescaled conformal Weyl spinor, the most important field in
these studies, from the F- into the NP-gauge, the factors $\rho^{\frac{1}{2}}$
and $\frac{1}{\rho}$ imply, due to cancellations and the integer spin of the
field, no singularities. The logarithmic term in $u$ does no harm in the
calculations either.

If the functions $u^p$ have been calculated and extend smoothly to
$I^+$, the calculation of the NP-gauge can in principle be extended
off ${\cal J}^+$ into $\tilde{M}$ to arbitrary order such that 
$\{u = const.\}$ will be the family of null hypersurfaces defining the
gauge. Furthermore, we can calculate an expansion of the NP fields on
${\cal J}^+$ in terms of $\rho$ and can extend the expansion suitably off
${\cal J}^+$ into $\tilde{M}$.

Since the functions $u^p$ on $I^+$ are given in terms of expressions
derived from the initial data on $\tilde{S} \cup I^0$, we obtain an expansion
of the NP fields on ${\cal J}^+$ near $I^+$ in terms of the initial data. In
particular, we can calculate expansions of the radiation field on ${\cal
J}^+$, the mass aspect, the Bondi-energy-momentum for a suitable family of
cuts of ${\cal J}^+$ etc. The resulting expressions contain non-trivial
information on the evolution of the gravitational field in time.  In
the case of data which are non-time-symmetric an analogous calculation
would allow us to relate non-trivial information of ${\cal J}^-$ to
information on ${\cal J}^+$. Of course, from the practical point of
view, the possibility to perform these calculations will be limited
by the sheer length of the latter, which will increase quickly with the order
of the calculation.

\section{The calculation of the NP-constants}

In \cite{NP:NPQ-letter}, \cite{NP:NPQ} Newman and Penrose discussed the
quantities
\begin{equation}
\label{NPconsts}
G_m =
\oint{_2\bar{Y}_{2,m}}\Psi^1_0\,{\rm{sin}}\vartheta\,d\vartheta\,d\varphi,
\,\,\,\,\,\,\,\,m = -2, -1, 0, 1, 2, 
\end{equation}
which are obtained by integration over a spherical cut ${\cal C}$ of 
${\cal J^+}$ where the fields entering the integral are given essentially in
the NP-gauge associated with the cut ${\cal C}$. Here $_2Y_{2,m}$ denote
spin-2 spherical harmonics, ${\rm{sin}}\vartheta\,d\vartheta\,d\varphi$ is
the surface element of ${\cal C}$, and $\Psi^1_0$ is the second coefficient
in the expansion $\Psi_0=\Psi^0_0r^{-5}+\Psi^1_0r^{-6}+O(r^{-7})$ of a
certain component of the ``physical'' conformal Weyl spinor in a certain
frame.  Here $r$ denotes a certain ``physical'' parameter along the
null generators of the hypersurfaces $\{u = const.\}$ with $r
\rightarrow \infty$ on ${\cal J}^+$.

As shown in \cite{NP:NPQ}, these quantities are "absolutely conserved" in the
sense that their values do not depend on the choice of the cut ${\cal C}$. 

Using the calculations indicated in the previous chapter, we can give the
``NP-constants'' (\ref{NPconsts}) for space-times satisfying our assumptions in
terms of quantities derived from the data $h$ on $S$. We express
the integrals (\ref{NPconsts}) in terms of the F-gauge, use their conservedness
to push ${\cal C}$ into $I^+$, express the result in terms of the functions
$u^p|_{I^+}$, and finally express the latter in turn in terms of the values of
$u^p$ on $I^0$. The resulting formula is of the form
\begin{equation}
\label{NPdatcon}
G_{m} = c_1\,m\,W_{2,2-m} - c_2\,R_{2,2-m} - 
\sum_{j + k = 2-m}\,c_{jk}\,W_{1,j}\,W_{1,k}, 
\end{equation}
with certain non-vanishing constants $c_i$, $c_{jk}$. The
coefficients $R_{2,2-m}$, $m = -2, -1, 0, 1, 2$, are derived from the
Ricci scalar $R$ of $h$ on the initial hypersurface. It has near 
$I^0 = \{\rho = 0, \tau = 0\}$ on $\{\tau = 0\}$ in the gauge employed in
\cite{FrKan} an expansion
\[
R = \frac{1}{2}\,(\sum_{m = 0}^4 R_{2,m}\,T_4\,^m\,_2)\,\rho^2 +
O(\rho^3).
\]
The coefficients $W_{2,2-m}$, $m = -2, -1, 0, 1, 2$, and $W_{1,j}$,$j = 0, 1,
2$, are obtained from $h$ as follows. The ``physical'' 3-metric $\tilde{h}$
is related to $h$ by a conformal factor $\Omega$ such that 
$h = \Omega^2\,\tilde{h}$. This factor has to satisfy the Lichnerowicz equation
$(\Delta_h - \frac{1}{8}\,R)(\Omega^{- \frac{1}{2}}) = 0$ on $\tilde{S}$
and certain fall-off conditions near space-like infinity. As a consequence it
can be given near $I^0$ on $\tilde{S} \cup I^0$ in the form 
$\Omega = \rho^2\,(U + \rho\,W)^{- 2}$ with some smooth functions $U$ and $W$. 
The function $U$ can be calculated locally from $h$. The function $W$
contains global information on the initial data. In the gauge used in 
\cite{FrKan} it has an expansion
\[
W = \frac{m}{2} + (\sum_{m = 0}^2 W_{1,m}\,T_2\,^m\,_1)\,\rho 
+ \frac{1}{2}\,(\sum_{m = 0}^4 W_{2,m}\,T_4\,^m\,_2)\,\rho^2 +
O(\rho^3),
\]
where $m$ denotes the ADM mass of $h$.

Newman and Penrose have derived for the NP-constants of static solutions an
expression of the form 
\[
mass\,\,\times\,\,quadrupole\,\,moment\,\,-\,\,(dipole\,\,moment)^2.
\]
Formula (\ref{NPdatcon}) is quite similar. If the ADM mass of $h$
vanishes, the coefficients $R_{2,m}$ vanish and in certain
representations of static solutions a factor $m$ can be extracted from
those coefficients. It should be noted, however, that, provided our
assumptions on the time evolution are satisfied, equation
(\ref{NPdatcon}) is valid for a class of time-symmetric solutions
which is much larger then the class of static solutions. In
particular, our solutions in general do not admit a time-like Killing
vector field and thus the evolution in time is no longer known
explicitly.

About the meaning of the NP-constants no consensus has been reached yet (cf.
the references given in \cite{FrKan}). Some light may be shed on this question
by analyzing the consequences of the coefficients entering formula
(\ref{NPdatcon}) on the structure of the solution.

\section{The calculation of $u^3$}

Since only second order quantities enter (\ref{NPdatcon}), the
quantities $u^p$, $p \le 2$, may appear sufficient to derive that
expression. However, the non-linear terms on the right hand side of
(\ref{NPdatcon}) are transported at third order and $u^3$ needs to be
calculated on $I^+$. As discussed in chapter 2, this is the first
order at which singularities may arise from the integral on the right
hand side of (\ref{gensol}). The proof that no such singularities occur
is quite involved because the expressions for $b$, derived recursively
from quantities of lower order, is getting more and more
complicated. However, apart from the fact that it would shed light on the
smoothness of the solution near $I^+$, the proof would be of interest
because it will give insight into the general form of $u^p$ and the
underlying properties of the field equations which imply this form. Clearly,
to show that the asymptotic regularity conditions are indeed sufficient
to exclude any further singularities, it is of interest to check this
for the first few orders.  Though the calculation is quite lengthy it
should give clues of how to give the general proof by a recursive
argument. Remarkably, in the case $p = 3$, we find by explicit
calculations that the integrand in (\ref{gensol}) has poles at $|\tau|
= \pm 1$ and also outside the interval $[-1, 1]$, that the integral
has poles and no logarithmic terms, but the final solution is a
polynomial in $\tau$. As a consequence, the frame, the connection
coefficients, and the components of the conformal Ricci tensor also
remain regular at the order $p = 4$. This may be considered as a first
indication that the asymptotic regularity conditions (\ref{asregcond})
suffice to ensure smooth extensibility of the $u^p$ to $I^{\pm}$. We
emphasize that only assumptions on the initial data have been used in
these calculations.

\section{Concluding remarks}

The discussion of \cite{FrKan} has shown that under suitable assumption it is
possible to analyze the behavior of Bondi systems in the setting of
\cite{Fr:space-like-infinity} to any desired degree of precision near
space-like infinity. Furthermore, the physical concepts studied on null
infinity can be calculated in principle to any order near $I^{\pm}$. We expect
that our results and the further exploration of space-like infinity will give
enough insight into the Einstein evolution near space-like infinity to
replace in the end the assumption on the smoothness of the fields on 
${\cal J}^{\pm}$ by the proof of existence results which imply the required
smoothness and which rely only on assumptions on the data. 

Another field in which the results of \cite{FrKan} should be of interest
is the numerical calculation of space-times. The setting of 
\cite{Fr:space-like-infinity} offers the only possibility to calculate 
numerically entire asymptotically flat space-times. Moreover, this can be
done by referring only to the asymptotically flat Cauchy data. This follows
from the finiteness of the coordinate representation of the regular finite
initial value problem. Thus there does not arise the need to introduce
time-like or null boundaries in the physical space-time to cope with the
unavoidable finiteness of numerical grids. Consequently, there never arise
questions about the meaning of the additional information fed into
space-time by prescribing boundary data. Finally, the conceptual (cf.
\cite{FrNag}) and numerical difficulties arising from time-like boundaries
can be avoided. The insights provided in \cite{FrKan} will be helpful to
perform controlled numerical calculations of space-times near space-like
infinity and to interprete the numerical results in terms of the physical
concepts introduced in the early studies of asymtptotically flat
space-times.   

\vspace*{0.25cm} \baselineskip=10pt{\small \noindent One of us (J.K.)
was supported by the research grant OTKA-D25135.}

\end{document}